\documentclass[aps,preprint,draft,tightenlines]{revtex4}

\begin{document}

\def\ap{\approx}

\def\beq{\begin{equation}}
\def\eeq{\end{equation}}
\def\haf{\frac{1}{2}}
\def\lpp{\lambda''}
\def\ccg{\cal G}
\def\slash#1{#1\!\!\!\!\!\!/}
\def\rpv{{\slash{R_p}\;}}
\def\u{{\cal U}}

\def\la{\mathrel{\mathpalette\fun <}}
\def\ga{\mathrelbe {\mathpalette\fun >}}
\def\fun#1#2{\lower3.6pt\vbox{\baselineskip0pt\lineskip.9pt
        \ialign{$\mathsurround=0pt#1\hfill##\hfil$\crcr#2\crcr\sim\crcr}}}

\renewcommand\({\left(}
\renewcommand\){\right)}
\renewcommand\[{\left[}
\renewcommand\]{\right]}

\newcommand\del{{\mbox {\boldmath $\nabla$}}}

\newcommand\eq[1]{Eq.~(\ref{#1})}
\newcommand\eqs[2]{Eqs.~(\ref{#1}) and (\ref{#2})}
\newcommand\eqss[3]{Eqs.~(\ref{#1}), (\ref{#2}) and (\ref{#3})}
\newcommand\eqsss[4]{Eqs.~(\ref{#1}), (\ref{#2}), (\ref{#3})
and (\ref{#4})}
\newcommand\eqssss[5]{Eqs.~(\ref{#1}), (\ref{#2}), (\ref{#3}),
(\ref{#4}) and (\ref{#5})}
\newcommand\eqst[2]{Eqs.~(\ref{#1})--(\ref{#2})}

\newcommand\pa{\partial}
\newcommand\pdif[2]{\frac{\pa #1}{\pa #2}}
                        
\newcommand\ee{\end{equation}}
\newcommand\be{\begin{equation}}
\newcommand\eea{\end{eqnarray}}
\newcommand\bea{\begin{eqnarray}}

\def\so{S_1}
\def\st{S_3}
\def\stb{\overline{S}_3}
\def\se{S_8}
\def\see{S'_8}
\def\sf{S_{15}}
\def\sfp{S'_{15}}

\def\vo{|S_1|^2}
\def\vt{|S_3|^2}
\def\vtb{|\overline{S}_3|^2}
\def\ve{|S_8|^2}
\def\vee{|S'_8|^2}
\def\vf{|S_{15}|^2}
\def\vfp{|S'_{15}|^2}

\newcommand\yr{\,\mbox{yr}}
\newcommand\sunit{\,\mbox{s}}
\newcommand\munit{\,\mbox{m}}
\newcommand\wunit{\,\mbox{W}}
\newcommand\Kunit{\,\mbox{K}}
\newcommand\muK{\,\mu\mbox{K}}

\newcommand\metres{\,\mbox{meters}}
\newcommand\mm{\,\mbox{mm}}
\newcommand\cm{\,\mbox{cm}}
\newcommand\km{\,\mbox{km}}
\newcommand\kg{\,\mbox{kg}}
\newcommand\TeV{\,\mbox{TeV}}
\newcommand\GeV{\,\mbox{GeV}}
\newcommand\MeV{\,\mbox{MeV}}
\newcommand\keV{\,\mbox{keV}}
\newcommand\eV{\,\mbox{eV}}
\newcommand\Mpc{\,\mbox{Mpc}}

\newcommand\msun{M_\odot}
\newcommand\mpl{M_{\rm P}}
\newcommand\MPl{M_{\rm P}}
\newcommand\Mpl{M_{\rm P}}
\newcommand\mpltil{\widetilde M_{\rm P}}
\newcommand\mf{M_{\rm f}}
\newcommand\mc{M_{\rm c}}
\newcommand\mgut{M_{\rm GUT}}
\newcommand\mstr{M_{\rm str}}
\newcommand\mpsis{|m_\chi^2|}
\newcommand\etapsi{\eta_\chi}
\newcommand\luv{\Lambda_{\rm UV}}
\newcommand\lf{\Lambda_{\rm f}}

\newcommand\lsim{\mathrel{\rlap{\lower4pt\hbox{\hskip1pt$\sim$}}
    \raise1pt\hbox{$<$}}}
\newcommand\gsim{\mathrel{\rlap{\lower4pt\hbox{\hskip1pt$\sim$}}
    \raise1pt\hbox{$>$}}}

\newcommand\diff{\mbox d}

\def\dbibitem#1{\bibitem{#1}\hspace{1cm}#1\hspace{1cm}}
\newcommand{\dlabel}[1]{\label{#1} \ \ \ \ \ \ \ \ #1\ \ \ \ \ \ \ \ }
\def\dcite#1{[#1]}

\def\dslash{\not{\hbox{\kern-2pt $\partial$}}}
\def\Dslash{\not{\hbox{\kern-4pt $D$}}}
\def\Oslash{\not{\hbox{\kern-4pt $O$}}}
\def\Qslash{\not{\hbox{\kern-4pt $Q$}}}
\def\pslash{\not{\hbox{\kern-2.3pt $p$}}}
\def\kslash{\not{\hbox{\kern-2.3pt $k$}}}
\def\qslash{\not{\hbox{\kern-2.3pt $q$}}}

 \newtoks\slashfraction
 \slashfraction={.13}
 \def\slash#1{\setbox0\hbox{$ #1 $}
 \setbox0\hbox to \the\slashfraction\wd0{\hss \box0}/\box0 }
 

\def\ee{\end{equation}}
\def\be{\begin{equation}}
\def\smallfrac#1#2{\hbox{${\scriptstyle#1} \over {\scriptstyle#2}$}}
\def\fourth{{\scriptstyle{1 \over 4}}}
\def\half{{\scriptstyle{1\over 2}}}
\def\st{\scriptstyle}
\def\sst{\scriptscriptstyle}
\def\mco{\multicolumn}
\def\epp{\epsilon'}
\def\vep{\varepsilon}
\def\ra{\rightarrow}
\def\ppg{\pi^+\pi^-\gamma}
\def\vp{{\bf p}}
\def\ko{K^0}
\def\kb{\bar{K^0}}
\def\al{\alpha}
\def\ab{\bar{\alpha}}

\def\calm{{\cal M}}
\def\calp{{\cal P}}
\def\calr{{\cal R}}
\def\calpr{{\calp_\calr}}

\newcommand\bfa{{\bf a}}
\newcommand\bfb{{\bf b}}
\newcommand\bfc{{\bf c}}
\newcommand\bfd{{\bf d}}
\newcommand\bfe{{\bf e}}
\newcommand\bff{{\bf f}}
\newcommand\bfg{{\bf g}}
\newcommand\bfh{{\bf h}}
\newcommand\bfi{{\bf i}}
\newcommand\bfj{{\bf j}}
\newcommand\bfk{{\bf k}}
\newcommand\bfl{{\bf l}}
\newcommand\bfm{{\bf m}}
\newcommand\bfn{{\bf n}}
\newcommand\bfo{{\bf o}}
\newcommand\bfp{{\bf p}}
\newcommand\bfq{{\bf q}}
\newcommand\bfr{{\bf r}}
\newcommand\bfs{{\bf s}}
\newcommand\bft{{\bf t}}
\newcommand\bfu{{\bf u}}
\newcommand\bfv{{\bf v}}
\newcommand\bfw{{\bf w}}
\newcommand\bfx{{\bf x}}
\newcommand\bfy{{\bf y}}
\newcommand\bfz{{\bf z}}

\newcommand\sub[1]{_{\rm #1}}
\newcommand\su[1]{^{\rm #1}}

\newcommand\supk{^{(K) }}
\newcommand\supf{^{(f) }}
\newcommand\supw{^{(W) }}
\newcommand\Tr{{\rm Tr}\,}

\newcommand\msinf{M\sub{inf}}
\newcommand\phicob{\phi\sub{COBE}}
\newcommand\delmult{\Delta V_{\chi\widetilde\chi{\rm f}}}
\newcommand\mgrav{m_{3/2}(t)}
\newcommand\mgravsq{m_{3/2}(t)}
\newcommand\mgravvac{m_{3/2}}

\newcommand\cpeak{\sqrt{\widetilde C_{\rm peak}}}
\newcommand\cpeako{\sqrt{\widetilde C_{\rm peak}^{(0)}}}
\newcommand\omb{\Omega\sub b}
\newcommand\ncobe{N\sub{COBE}}
\newcommand\vev[1]{\langle{#1}\rangle}


\title{Cosmological constraints on 
a Peccei-Quinn flatino as the lightest
supersymmetric particle}
 
\author{Eung Jin Chun$^{a,b}$, Hang Bae Kim$^a$ and David H. Lyth$^a$}

\affiliation{
$^a$Department of Physics, Lancaster University, Lancaster LA1 4YB, UK\\
$^b$Korea Institute for Advanced Study, Seoul 130-012, Korea}

\date{\today}

\begin{abstract}
In an interesting class of models, non-renormalizable terms of the
superpotential are responsible for the spontanteous breaking of 
Peccei-Quinn (PQ) symmetry as well as the generation of the $\mu$ term. 
The flaton fields which break PQ symmetry are
accompanied by flatinos, and the lightest flatino (the LSP) can be
the stable, while the decay of the lightest  neutralino  (the NLSP) 
might be visible at colliders with a low axion scale.
We examine the cosmology of these models 
involving    thermal inflation
just after the PQ phase transition.
The branching ratio of flatons  into
axions must be small so as not to interfere with
nucleosynthesis, and  flatons must  not decay into the LSP
or it will be over-abundant.
We explore a simple model, 
with  light flatons which can decay into $Z$ or $W$ bosons, 
or into a light Higgs ($h^0$) plus a $Z$ boson, 
to show that such features can be realized in a wide range 
of parameter space.
The mass of the NLSP can be as low as $(m_{h^0}+m_Z)/2$, 
with an axion scale of order $10^{10}$ GeV and
a final reheat temperature typically of order $10\GeV£$.
Then, the flatino LSP is a good dark matter
candidate because the reheat temperature can be high enough to allow
its production from the decay of the thermalized LSP, while  low enough 
to prevent its overproduction from the decay of sfermions.
\end{abstract}

\pacs{PACS number(s): 11.30.Hv, 12.15.Ff, 12.60.Jv, 14.60.Pq }

\maketitle


\section{Introduction}

In supersymmetric extensions of the Standard Model, both the 
 $\mu$ problem and the strong CP problem can be  resolved
through  a simple extension of the Higgs sector, which implements
spontaneously broken 
 Peccei-Quinn (PQ) symmetry \cite{pq}.  
Suppose that the Higgs bilinear term in
the minimal supersymmetric standard model (MSSM) superpotential takes
the non-renormalizable form; 
\begin{equation} \label{muterm}
 W_{\mbox{\scriptsize$\mu$-term}}=  
h{ \prod_i\phi_i^{l_i} \over M_P^{p-1}} H_1 H_2
\end{equation}
where $M_P\simeq 2.4\times 10^{18}$ GeV is the reduced Planck mass,
$p=\sum_i l_i \geq 2$, and $W_{\mbox{\scriptsize$\mu$-term}}$ respects a 
PQ symmetry.  Then, 
nonzero vacuum expectations values
(vevs) $\langle \phi_i \rangle$ can
spontaneously  break
 PQ symmetry,  providing  the axion solution to the 
strong CP problem \cite{areview},  while   generating the right order of
 magnitude for the 
 $\mu$ parameter \cite{muprob}; 
\begin{equation} \label{muis}
\mu=h {\prod_i\langle \phi_i\rangle^{l_i} \over M_P^{p-1}} \,.
\end{equation} 
In the case of a polynomial of the second order ($p=2$), 
with $h\sim 1$, 
this gives  $\mu \sim F\sub a^2/M_P \sim 100$ GeV if  the axion scale
$F\sub a \sim \langle \phi_i \rangle$ is of order $10^{10}$ GeV, its lowest
feasible value.
With a higher power $p>2$, one would need a larger value of $F\sub a$ 
to get the right size of $\mu$.

{}From now on, we consider the case of supergravity with gravity-mediated
supersymmetry breaking  \cite{nilles}. There are two ways of
 spontaneously breaking
 PQ symmetry. One is to introduce a renormalizable tree-level
superpotential. In this case,  the particles corresponding to the 
fields  $\phi_i$, and their spin $1/2$ partners, all have 
 mass of order $F\sub a$ except for a supermultiplet comprising
 the  axion (the pseudo-Goldstone boson of the PQ symmetry), the
spin zero saxion, and the spin $1/2$ axino.  
The mass of the axino in this case
is  generically of order $100$ GeV,
but it can be as small as O(keV)
 \cite{goto,andre}. If it is the lightest supersymmetric particle
 (LSP), there follow some interesting consequences in cosmology 
and collider physics.  If the axino mass is
 $\gtrsim 10$ GeV, it 
can be the  cold dark matter of the universe if its (non-thermal)
production comes mostly  from the decay of the next-to-lightest
supersymmetric particle (NLSP)  \cite{covi}.
For this to happen, of course, the primordial LSP relic density
has to be diluted away, which would be the case when the reheat 
temperature after inflation is sufficiently low \cite{chun2}.
(In contrast,  the axino mass is required to be less than a few keV if
its primordial abundance is not washed out \cite{raja}.)

The  alternative  to a renormalizable superpotential
is a non-renormalizable  superpotential with soft
supersymmetry breaking, leading to a flaton model of
of  PQ symmetry breaking
\cite{mox}--\cite{denis}. (A flaton \cite{yam85,lyth}
is a  scalar field with vev much bigger than its mass, corresponding
to a very  flat potential.)
For simplicity, the superpotential is supposed to be dominated by
a single term,
\begin{equation} \label{PQsector}
 W_{PQ}= {f \over M_P^{n-3}} \prod_i \phi_i^{k_i}
\end{equation}
where $n=\sum_i k_i \geq 4$.
If the soft mass-squared of one or more of the singlet fields
 $\phi_i$ is  negative, e.g., through a radiative mechanism \cite{msy}, 
there exists always a nontrivial global minimum.
The  $\phi_i$ with  nonzero vevs are then  flatons.
The  axion scale is
$F\sub a\sim \langle \phi_i \rangle \sim (m_0 M_P^{n-3})^{1/n-2}$,
which is naturally of the right size.
For instance, for $n=4$,  $f\sim 1$ and
 $m_0\sim 100$ GeV, the vacuum expectation value  is 
of order $10^{10}$ GeV, but it can be bigger if $n$ is bigger.

In a flaton model, the masses of the 
 particles corresponding to the 
fields  $\phi_i$, and their spin $1/2$ partners, all have 
 mass of order $100$ GeV, with the sole exception of the axion.
The saxion and axino, defined as the superpartners of the axion,
have no special significance and are in general not mass eigenstates.
Instead, with $N\geq 2$ fields breaking PQ symmetry, there are $2N-1$
spin zero flaton particles, and $N$ spin $1/2$ flatinos. These play
the same role as, respectively, the saxion and the axino of renormalizable
models. In particular, the lightest flatino may be the LSP and hence a dark
matter candidate.

With either an axino or a flatino LSP, the
 NLSP, which
is typically the lightest neutralino in the MSSM, decays to the LSP
with the rate proportional to $1/F\sub a^2$.
An  interesting implication for collider experiments has been 
pointed out  in Ref.~\cite{martin}.
Although the  NLSP decays are very weak,
and thus the corresponding decay lengths can be much larger than
a collider size, 
there is an opportunity to observe these decays in future colliders
with a large number of supersymmetric events so as to provide 
direct information about physics at very high energy scale.

In this  paper, we  study the cosmology of models with a
 flatino  LSP, identifying  various bounds on the parameters which are required
for a viable cosmology. The paper
 may regarded as a sequel to \cite{denis}, in which all of the flatinos
were  supposed to be unstable.
We shall focus particularly on the case 
of a low axion scale, $F\sub a \sim 10^{10}$ GeV, which increases
the possibility of collider signatures.  

\section{Cosmology of flatino LSP}

Flaton models lead to very interesting cosmological effects.
On the reasonable assumption that the flaton fields are nonzero
in the early Universe, they lead to 
thermal inflation which can dilute away all the unwanted relics 
\cite{lyth,laza2}.  In a flaton model of PQ symmetry breaking, 
some restrictions have to be put on the model in order not to
overproduce unwanted relics again from flaton decay.
As a (scalar) flaton field $\varphi$ has always the decay channel
$\varphi \to aa$ with a rate $\Gamma_{\varphi \to aa}\sim 
m_\varphi^3/32\pi F\sub a^2$ \cite{andre},
flaton decay may produce too  many (unthermalized) axions and upset
 standard nucleosynthesis.  In the scenario of the flatino LSP
under consideration, it is also required that the LSP is not
overproduced from flaton decay.
In general, a flaton $\varphi$  
can decay into ordinary particles and their superpartners $X$,
the axion $a$ and the flatino LSP $\tilde{F}_1$.
After flaton decay, the energy densities  are given
by $\rho_X=B_X \rho_\varphi=(\pi^2/30) g_{RH} T\sub{RH}^4$, 
$\rho_a=B_a \rho_\varphi$ and 
$\rho_{\tilde F_1}=B_{\tilde F_1}(m_{\tilde F_1}/m_\varphi) \rho_{\varphi}$ 
where $\rho_\varphi$ is the
energy density of the flatons before they decay  and the $B$'s 
are essentially branching ratios of the flaton decay.
The precise relation between $B_a$ and the flaton branching ratios
is given in \cite{denis}.
As is well-known, nucleosynthesis (NS) puts a constraint on 
the extra amount of relativistic energy density, which is
conveniently given in terms of the  equivalently  number 
of neutrino species, $\delta N_\nu$.
At present \cite{nsbound}, the bound is something like
$\delta N_\nu< 0.3$ in the favored `low deuterium' scenario.
This can be translated to  upper limits on 
$B_a$ and  $B_{\tilde F_1}$.
Applying $(\rho_a/\rho_\nu)_{NS}< \delta N_\nu$ for the energy 
densities at the time of nucleosynthesis, one finds \cite{choi,denis}
\begin{equation} \label{Babound}
 B_a < {7\over 43} \left(g_{RH} \over 43/4\right)^{1/3} \delta N_\nu \,.
\end{equation}
In the similar way, the energy density of the stable and massive LSP
$\tilde F_1$ has to be constrained; 
$(\rho_{\tilde F_1}/\rho_\nu)_{NS} < \delta N_\nu$, 
which leads to  
\begin{equation} \label{BtF1}
 B_{\tilde F_1} < {7/4 \over  g_{RH}} 
         { m_{\varphi} \over m_{\tilde F_1}}
         { T_{NS} \over T\sub{RH} } \delta N_\nu \,.
\end{equation}

To get a rough estimate of the reheat temperature, let us
take the rate of the decay $\varphi\rightarrow X$ to be 
$\Gamma_X \sim \Gamma_a/B_a$ resulting in 
\begin{equation} \label{TRH}
 T\sub{RH} \approx 1.2 g_{RH}^{-1/4} \sqrt{M_P \Gamma_X}
        \sim 19 \,{\rm GeV}
         \left(m_\varphi \over 100\,{\rm GeV} \right)^{3/2}
        \left(10^{10}\, {\rm GeV} \over  F\sub a \right)
\end{equation}
where $g_{RH}\sim 100$ is the effective number of particle degrees 
at $T\sub{RH}$ and we took the numerical factor $B_a=0.1$.  

For $T\sub{RH} \sim 10$ GeV, we get
$B_{\tilde F_1} \lesssim 10^{-4}$ with $B_a \lesssim 0.1$.
A more stringent limit on $B_{\tilde F_1}$ comes from the flatino 
contribution to the present energy density.
That is, in order to avoid the overclosure by the LSP,
$(\rho_{\tilde F_1}/\rho_c)_0 <1$, one gets 
\begin{equation} \label{BtFbound}
 B_{\tilde F_1} \lesssim {4\over3} { m_{\varphi} \over m_{\tilde F_1}}
              \left(3.55\,{\rm eV} \over T\sub{RH}\) B_X \,.
\end{equation}
Taking $T\sub{RH} \sim 10$ GeV,  
one obtains  $B_{\tilde F_1} \lesssim 10^{-10}$.
This is a very strong constraint.  
It appears impossible to get such a small number without relying on 
a severe fine-tuning of parameters in the model once flaton decay into
flatino  is allowed kinematically.  Therefore, we have to forbid
the decay of flatons into the flatino LSP imposing 
\begin{equation} \label{mass1}
  m_\varphi < 2 m_{\tilde F_1} \,.
\end{equation}
Of course, the condition (\ref{BtFbound}) can be invalidated 
in the case that the flatino  is heavy enough to decay into ordinary 
particles, e.g., a neutralino and a light Higgs boson \cite{denis}
which is opposite to our consideration.

Forbidding the decay $\varphi\to \tilde F_1$ kinematically, we now have to
fulfill the condition (\ref{Babound}).  
Important thermalizable decay modes of a flaton 
to achieve $B_a \ll B_X \approx 1$ include
the decays into two top quarks, two light stops, 
two light Higgs bosons $h^0$, two $Z$/$W$ gauge bosons 
and a Z boson plus a Higgs boson. 
In this paper, we take the flatons to be as light as possible, so as to allow
the lightest possible masses
for the flatino  LSP and for the NLSP. We therefore assume that 
only the  last two decay modes are kinematically allowed.
The decay mode into Higgs bosons comes from direct couplings of Higgs bosons
and flatons  given by Eq.~(\ref{muterm}) and has been considered
in Ref.~\cite{denis} in the context of an  ordinary 
neutralino LSP.  The other modes listed above
come from   mixing between Higgs 
bosons and flatons induced by the same $\mu$ term interaction (\ref{muterm}),
which is the 
 bosonic counterparts of mixing between neutralinos 
and  flatinos  which has been worked out in Ref.~\cite{martin}.
(The qualitative features of the decay modes into top quarks or 
stops have  also been considered in Ref.~\cite{choi}.)

Under the condition of no CP violation, scalar flatons (denoted by $F$)
can decay into two gauge bosons, e.g.,  $F\to WW$, and pseudoscalars 
(denoted by $F'$) into a Z boson and a scalar Higgs, $F'\to h^0 Z$, 
as we describe below.
The former mode gives a  lower limit on the scalar flaton masses, 
$2m_W <m_F$. Combining it
 with Eq.~(\ref{mass1}) we find
\begin{equation} \label{mass2}
 2 m_W <  m_F <2 m_{\tilde F_1} < 2 m_{\tilde{\chi}_1^0} 
\end{equation}
where $\tilde{\chi}_1^0$ is the NLSP neutralino.
(When $m_F < 2m_W$ for instance, flatons can of course decay into 
three ordinary light fermions mediating a sfermion.  
But the corresponding decay rates have a large 
phase space suppression which makes it hard to dominate over the
decay mode $F\to aa$.) 
Concerning  pseudoscalar flatons, there are more possibilities.
When $m_{F'}>m_F$, the decay mode $F'\to a F$ is open and 
has to be suppressed against the mode $F'\to h^0 Z$.
This requires
\begin{equation} \label{mass3}
m_{h^0} + m_Z < m_{F'} < 2m_{\tilde F_1} < 2m_{\tilde{\chi}_1^0} \,.
\label{10}
\end{equation}
On the other hand, if $m_{F'} < m_F$, $F'$ has no two body decay mode
into an axion,  and  thus $F'$ can be so light as to have the
decay modes into light fermions; $F' \to f \bar{f}(Z)$ or $f\bar{f} h^0$.
In this case, the corresponding decay rate
has a large Yukawa-coupling or phase-space suppression factor 
leading to lower reheat temperature than in Eq.~(\ref{TRH}), 
and only the bound (9) is applied. 

\medskip

Without resorting to a specific model,
let us describe how the flaton couplings to gauge bosons
arise.  If we assume no CP violation, scalar flatons  mix with the  two
CP-even Higgs bosons $h^0, H^0$ and pseudoscalar flatons
mix with the CP-odd Higgs  boson $A$, through the $\mu$-term potential
as in Eq.~(\ref{muterm}).  Quantifying small mixtures of
flatons in Higgs bosons by $\varepsilon_{h^0F}$, $\varepsilon_{H^0F}$
and $\varepsilon_{AF'}$ in a self-explaining notation, 
it is straightforward to write down the flaton-Higgs couplings 
from the MSSM Lagrangian: 
\begin{eqnarray} \label{HFcoupling}
 {\cal L}&=&-gm_W \varepsilon_{VVF} 
      [FW^+_\mu W_-^\mu +{1\over c_W^2}  FZ_\mu Z^\mu ]
      \nonumber\\
 &&-{g\over c_W}\varepsilon_{AF'}
    [\cos(\alpha-\beta) h^0 +\sin(\alpha-\beta) H^0] 
    \overline{\partial_\mu}F' Z^\mu
\end{eqnarray}
where 
$\varepsilon_{VVF}\equiv \sin(\alpha-\beta) \varepsilon_{h^0F}
           -\cos(\alpha-\beta) \varepsilon_{H^0F}$, 
$\alpha$ is the diagonalization angle of CP-even Higgses, 
$\beta$ is defined by $\tan\beta=\langle H_2 \rangle/\langle H_1 \rangle$
and $c_W=\cos\theta_W$ with $\theta_W$ being the weak mixing angle.

Let us now consider the flatino--neutralino mixing \cite{martin}
which arises at tree level due to the $\mu$ term (\ref{muterm}).
For typical parameters, this 
 mixing  will give the main contribution to the flatino 
interactions with ordinary particles and their superpartners,
 dominating  the
one-loop processes \cite{covi} 
which  are otherwise responsible for these interactions.
The relevant interactions for our purpose are 
the decay of the  NLSP to  the flatino LSP, and the decays of sfermions
to the flatino LSP.

The term (\ref{muterm}) leads to 
the mass matrix that mixes  $\tilde{F}_i$ and ($\tilde{B}$, $\tilde{W}_3$,
$\tilde{H}_1$, $\tilde{H}_2$) as follows;
\begin{equation} \label{ftinomix}
 \pmatrix{ 0 & 0 & \mu s_\beta \delta_i & \mu c_\beta \delta_i \cr}
\,,
\end{equation}
where $c_\beta\equiv \cos\beta$,
$s_\beta\equiv \sin\beta$, and 
the coefficients $\delta_i \sim v/F\sub a$ depend on the
specific form of the superpotential (\ref{muterm}).
Here $v=264$ GeV is the Higgs vev.
Let $N$ be the diagonalization matrix of the MSSM neutralino mass matrix
$M_N$: $\tilde{M}_N= N^T M_N N$ where $\tilde{M}_N$ is the diagonalized
neutralino mass matrix for the eigenstate $\tilde{\chi}^0_j$.
Then, further diagonalizing  (\ref{ftinomix}) can be done
by the small mixing elements;
\begin{equation}  \label{ftinodiag}
 \varepsilon_{\tilde{\chi}_j^0 \tilde{F}_i} = \delta_i 
 {\mu (s_\beta N_{3j}+c_\beta N_{4j}) \over 
    m_{\tilde{\chi}_j^0}- m_{\tilde{F}_i} }
\end{equation}
which describe the small content of the flatino $\tilde{F}_i$ in 
the neutralino $\tilde{\chi}_j^0$.   Note that we can have an
enhancement in the mixing (\ref{ftinodiag}), 
$\varepsilon_{\tilde{\chi}_j^0 \tilde{F}_i} \gtrsim 10 v/F_a$, 
when $  m_{\tilde{\chi}_j^0}- m_{\tilde{F}_i} = O(10)$ GeV.

Flaton-Higgs mixing parameters and flaton/flatino masses are dependent 
on specific forms of the terms (\ref{muterm}) and (\ref{PQsector}).
As we discussed, in order for the flatino LSP scenario to 
be consistent with cosmological considerations,
the model should fulfill the mass relations (\ref{mass1}), (\ref{mass2})
and (\ref{mass3}), and 
have sufficiently large mixing elements $\varepsilon$ in 
Eq.~(\ref{HFcoupling}) to suppress the axion decay modes.
However, note that mixing between the flaton sector 
and the MSSM sector is determined by the term (\ref{muterm}) and  
flaton/flatino mass spectrum by the term (\ref{PQsector}) 
and its soft-breaking term.  Therefore, arrangement for the mass 
relations (\ref{mass1},\ref{mass2},\ref{mass3}) and large mixing elements 
$\varepsilon$ can be done by independent parameters in two separate 
sectors, which implies that such an arrangement can be easily 
achieved in generic flaton models.  

\medskip

As  the flatino LSP has  a mass in the 100 GeV region,
its  slight  regeneration after the thermal inflation
can provide a sizable contribution to the matter density at present.
There are two important sources of the flatino regeneration.
The first is  the neutralino decay \cite{covi} 
after its decoupling from the thermal bath.
In the flaton models, the NLSP neutralino can decay 
into $Z(h^0) \tilde F_1$ or $f \bar{f} \tilde F_1$ within 0.01 sec
\cite{martin} and thus without affecting nucleosynthesis.
Then, the ratio of the flatino LSP density to the critical density is
given by 
\begin{equation} \label{fromdecay}
 \Omega_{\tilde F_1} = { m_{\tilde F_1} \over m_{\tilde{\chi}_1^0} }
 \Omega_{\tilde{\chi}_1^0}  \,.
\end{equation}
For our case, $m_{\tilde F_1}/m_{\tilde{\chi}_1^0} = {\cal O}(1)$ and 
thus we can have $ \Omega_{\tilde F_1} \sim\Omega_{\tilde{\chi}_1^0} 
\sim 0.1-1 $ which is valid in a wide range of the MSSM parameter space
\cite{drees}.  For this mechanism to work, the reheat temperature
(\ref{TRH}) after thermal inflation should be larger than the neutralino
decoupling temperature $\sim m_{\tilde{\chi}_1^0}/20$.  
As can be seen from Eq.~(\ref{TRH}), 
this condition holds for low $F\sub a \sim 10^{10}-10^{11}$ GeV which 
increases collider signals as well.

A potentially more important source is  thermal regeneration \cite{covi2}.
Since our reheat temperature is typically below 100 GeV, decay processes
dominate over scattering processes for the thermal regeneration.
To get a qualitative calculation of the flatino population, let us
take a sfermion--fermion--flatino interaction arising from the 
neutralino--flatino mixing, 
\begin{equation}
 {\cal L} =  -g \varepsilon_{\tilde{f}f\tilde F_1} 
              \tilde{f} f \tilde F_1 + h.c.
\end{equation}
where $\varepsilon_{\tilde{f}f\tilde F_1} \propto v/F\sub a$ contains the 
factors from gauge quantum numbers and neutralino--flatino mixing and
$\tilde{f}$/$f$ denotes a left-handed or right-handed 
sfermion/fermion field.
The thermally regenerated flatino population is given by
$\Omega_{\tilde F_1} h^2= m_{\tilde F_1} Y_{\tilde F_1}/3.55\,{\rm eV}$
with the factor \cite{covi2,hbkim} :
\begin{equation}
  Y_{\tilde F_1} \approx 2\times10^{-5} \frac{M_P \Gamma_{\tilde f}}
{   T\sub{RH}^2 }  F(x_{\tilde f})
\end{equation}
where $\Gamma_{\tilde f}= g^2 \varepsilon_{\tilde{f}f\tilde F_1}^2 
m_{\tilde{f}} /8\pi$ is the decay rate and $F(x_{\tilde{f}})$ is the Boltzmann 
suppression factor as a function of $x_{\tilde{f}} = m_{\tilde{f}}/T\sub{RH}$ 
given by
$$  F(x_{\tilde{f}}) \equiv {1\over x_{\tilde{f}}^2}
   \int_{x_{\tilde{f}}}^\infty {d\!x \over e^x-1} \left[\left( 
    {\pi\over2} -\tan^{-1}{x \over \sqrt{x^2-x^2_{\tilde{f}}}}
    \right)x^4
    + x_{\tilde{f}}\left(x^2-2x_{\tilde{f}}  \sqrt{x^2-x^2_{\tilde{f}}} 
     \right) \right] \,. $$
We have defined the function $F(x_{\tilde{f}})$ 
to remove the dependence on the 
axion scale $F\sub a$ for the remaining factor;
$M_P\Gamma_{\tilde{f}}/T\sub{RH}^2 \propto \Gamma_{\tilde{f}}/\Gamma_\varphi$
where $\Gamma_\varphi$ is the flaton decay rate.
To a very good approximation for the range of $x\gtrsim 5$, 
the function $F(x)$ can be expressed in terms of an analytic function;
$F(x) \approx 36 e^{-0.98x}$ 
giving
\begin{equation}
 \Omega_{\tilde F_1} h^2 \approx   5\times10^4
      \left( m_{\tilde F_1}\over m_{\tilde{f}} \right)
      \left( \varepsilon_{\tilde{f}f\tilde F_1} \over 10^{-8} \right)^2
      x_{\tilde{f}}^2 e^{-0.98 x_{\tilde{f}}} \,.
\label{17}
\end{equation}
{}From the above equation, one finds that the requirement
$ \Omega_{\tilde F_1} h^2 \lesssim 1$ gives $x_{\tilde{f}} \gtrsim 17$
with $m_{\tilde F_1}= m_{\tilde{f}}$ and 
$\varepsilon_{\tilde{f}f\tilde F_1} = 10^{-8} $.
This consideration gives some meaningful bounds on sfermions masses,
$m_{\tilde{f}} > 17 T\sub{RH}$ given the reheat temperature (\ref{TRH}),
or vice versa.
The thermally regenerated population of the flatino LSP can
also be the cold dark matter which, however, needs a fine
arrangement for the sfermion mass and reheat temperature 
as $ \Omega_{\tilde F_1} h^2$ is a very sensitive function of 
$x_{\tilde{f}}$.
Note that the thermal regeneration becomes easily negligible 
as $F\sub a$ becomes large.  For instance, for $F\sub a=10^{11}$ GeV, 
one has $T\sub{RH} \sim 2$ GeV (\ref{TRH}) and thus 
$ x_{\tilde{f}} \gtrsim  50$ for $m_{\tilde{f}} > 100$ GeV.

\section{A Model}

To illustrate our discussion more explicitly, we
consider the minimal case of two fields and $n=4$,
 which has been analyzed in Ref.~\cite{denis}.
This model contains two scalar flatons,  one pseudoscalar flaton, 
and two flatinos.
The flaton superpotential  is 
\begin{equation} \label{WPQ}
 W_{PQ} = {f \over M_P} P^3 Q 
\end{equation}
with $U(1)_{PQ}$ charges of $P$ and $Q$ being $1$ and $-3$, respectively.
Analyzing the scalar potential including soft supersymmetry breaking terms
\begin{equation} \label{VPQ}
V_{soft} = {f A_f \over M_P} P^3 Q + h.c. \,,
\end{equation}
one finds  scalar flatons $F_{2,1}$ with  masses-squared
\begin{equation}
 m^2_{F_2, F_1} = {1 \over 2} f^2 \mu_0^2 
   \left( 3(12 - \xi) + x^2(12+ \xi)
   \pm |12-\xi| \sqrt{x^4+42x^2+9} \right)
\end{equation}
and a pseudoscalar flaton $F'$ with mass-squared
\begin{equation}
 m^2_{F'}= f^2 \mu_0^2 \xi (x^2+9)
\end{equation}
where the parameters are defined by
$$ x \equiv {\langle P \rangle  \over \langle Q \rangle}\,,\quad
   \xi \equiv -{A_f \over f\mu_0} > 0\,,\quad  
   \mu_0 \equiv  {\langle P \rangle  \langle Q \rangle \over 2M_P}\,. $$
In the above, the flatons $P,Q$ are expanded as 
\begin{eqnarray}
 P &=& {1\over \sqrt{2}} \left( \langle P \rangle + P' 
       - 3i{\langle Q \rangle \over F\sub a} F' \right) \nonumber\\
 Q &=& {1\over \sqrt{2}} \left( \langle Q \rangle + Q' 
       - i{\langle P \rangle \over F\sub a} F' \right)
\end{eqnarray}
where $F\sub a^2 \equiv \langle P \rangle^2 + 9 \langle Q \rangle^2$.
The diagonalization matrix of the scalar flatons $P'$ and $Q'$ is
\begin{equation}
\pmatrix{ P' \cr Q' \cr} = 
\pmatrix{ \cos\varphi & -\sin\varphi \cr
          \sin\varphi & \cos\varphi \cr} 
\pmatrix{ F_2 \cr F_1 \cr}
\end{equation}
where the mixing angle $\varphi$ is determined by
\begin{eqnarray}
 \cos2\varphi &=& sgn(12-\xi) {x^2+3 \over \sqrt{x^4+42x^2+9} } \nonumber \\
 \sin2\varphi &=& sgn(12-\xi) { 6x \over \sqrt{x^4+42x^2+9} }  \,.
\end{eqnarray}
{}From the superpotential (\ref{WPQ}), one also finds the flatino
masses
\begin{equation}
 m_{\tilde F_2, \tilde F_1} = 
 3 f \mu_0 [\sqrt{x^2+1} \pm 1]
\end{equation}
The light flatino $\tilde F_1$ is supposed to be the 
 LSP.
The rotation matrix from the flavor states $\tilde{P},\tilde{Q}$ to
the mass eigenstates $\tilde{F}_2, \tilde F_1$ is given by
\begin{equation}
\pmatrix{ \tilde{P} \cr \tilde{Q} \cr} = 
\pmatrix{ \cos\tilde{\varphi} & -\sin\tilde{\varphi} \cr
          \sin\tilde{\varphi} & \cos\tilde{\varphi} \cr} 
\pmatrix{ \tilde F_1 \cr \tilde{F}_2 \cr}
\end{equation}
with the mixing angle $\tilde{\varphi}$ satisfying 
\begin{equation}
 \cos2\tilde{\varphi} = - {1\over \sqrt{x^2+1} } \,,\quad
 \sin2\tilde{\varphi} = - { x\over \sqrt{x^2+1} }  \,.
\end{equation}
As all the masses contain the overall factor $|f\mu_0|$, 
it is now straightforward to find the region of the parameters
$x$ and $\xi$ satisfying the desired mass relations.
It turns out that $m_{F_{1,2}},\; m_{F'} < 2 m_{\tilde F_1}$ 
requires
\begin{eqnarray} \label{AB}
(A)&\quad& \xi <12; \quad x>2.6,\quad \xi_{a,b} < \xi <\xi_c  \nonumber\\
(B)&\quad& \xi >12; \quad x>3.5,\quad 12 < \xi <\xi_c  \\
{\rm with} &\quad&
\xi_a =12{\sqrt{x^4+42x^2+9}-x^2-3 \over\sqrt{x^4+42x^2+9}+x^2-3} \nonumber\\
&&\xi_b =12{\sqrt{x^4+42x^2+9}+12\sqrt{x^2+1}-5x^2-9 \over
         \sqrt{x^4+42x^2+9}+x^2-3} \nonumber\\
&&\xi_c =36{x^2+2-2\sqrt{x^2+1} \over x^2+9} \nonumber
\end{eqnarray}
We always have $m_{F_1} < m_{F'}, m_{F_2}$ in the regions (A,B) and 
also $m_{F_2} < m_{F'}$ in the region (B). 
The requirement (\ref{mass3}) has to be fulfilled  in both regions.

The important interactions of flatons (flatinos) and MSSM fields 
comes from the mixings between flatons (flatinos) and Higgs bosons 
(neutralinos) due to the $\mu$-term superpotential of our choice
\begin{equation} \label{WmuPQ}
 W_{\mbox{\scriptsize$\mu$-term}} = h {PQ \over  M_P} H_1 H_2 
\end{equation}
with $U(1)_{PQ}$ charge of $H_1H_2$ being $+2$.
Note that $\mu = h \langle PQ\rangle /2M_P = h \mu_0$.
First of all, Eq.~(\ref{WmuPQ}) determines 
the flatino--Higgsino  mixing elements 
$\delta_i$ defined in Eq.~(\ref{ftinomix}) as follows:
\begin{equation} 
\delta_1={v\over F\sub a} {\sqrt{x^2+9} \over x} 
     (c_{\tilde{\varphi}}+x s_{\tilde{\varphi}}) \,,\quad
\delta_2={v\over F\sub a} {\sqrt{x^2+9} \over x} 
     (-s_{\tilde{\varphi}}+x c_{\tilde{\varphi}}) \,,
\end{equation} 
with $c_{\tilde\varphi}\equiv\cos _{\tilde\varphi}$ and $s_{\tilde\varphi} 
\equiv \sin_{\tilde\varphi}$.
Including the soft term of (\ref{WmuPQ}), we get the Higgs-flaton 
scalar potential,
\begin{eqnarray} \label{VHF}
V &=& \left| H_1 \right|^2
\left( m_{H_1}^2 + \left| h{ P Q \over M_P} \right|^2 \right)
+ \left| H_2 \right|^2
\left( m_{H_2}^2 + \left| h{PQ \over M_P} \right|^2 \right)
 \nonumber  \\
 &&
+ \left\{ h H_1 H_2 \left( A_h{P Q \over M_P} +
   3f^*{ P^{*2}| Q|^2 \over M_P^2} +
   f^*{ P^{*2}|P|^2 \over M_P^2} \right) +
 \mbox{c.c.} \right\} \\
&&
 + {1\over8} (g^2+ {g'}^2) \left( |H_1|^2 - |H_2|^2 \right)^2  \nonumber \,.
\end{eqnarray} 
We note that the CP-odd Higgs boson has mass-squared
\begin{equation}
 m_A^2= -{2 \over \sin2\beta}\mu^2
        \left( {A_h\over \mu} + {f\over h}(x^2+3) \right)\,.
\end{equation}
It is then straightforward to find the following flaton--Higgs mixing:
\begin{eqnarray}
\varepsilon_{h^0 F_1} &=& {v \over F\sub a} {\sqrt{x^2+9} \over x} 
     {\mu^2 \over m_{h^0}^2-m^2_{F_1} } \{ 
 -\sin(\alpha-\beta)[2s_\varphi-2xc_\varphi]   \\
&& 
 +\cos(\alpha-\beta)[{f\over h} (4x^2 s_\varphi + 6 s_\varphi - 6xc_\varphi) 
                  + {A_h \over \mu} (s_\varphi -xc_\varphi)] \} 
       \nonumber \\
\varepsilon_{H^0 F_1} &=& {v \over F\sub a} {\sqrt{x^2+9} \over x} 
        {\mu^2 \over m_{H^0}^2-m^2_{F_1} } \{ 
 \cos(\alpha+\beta)[2s_\varphi-2xc_\varphi] \\
&&
 + \sin(\alpha+\beta)[{f\over h} (4x^2 s_\varphi + 6 s_\varphi - 6xc_\varphi) 
                  + {A_h \over \mu} (s_\varphi -xc_\varphi)] \} 
       \nonumber \\  
\varepsilon_{h^0 F_2} &=& {v \over F\sub a} {\sqrt{x^2+9} \over x} 
        {\mu^2 \over m_{h^0}^2-m^2_{F_2} } \{
 \sin(\alpha-\beta)[2c_\varphi+2xs_\varphi]  \\
&& 
  -\cos(\alpha-\beta)[{f\over h} (4x^2 c_\varphi + 6 c_\varphi + 6xs_\varphi) 
                  + {A_h \over \mu} (c_\varphi +xs_\varphi)] \}
      \nonumber  \\  
\varepsilon_{H^0 F_2} &=& {v \over F\sub a} {\sqrt{x^2+9} \over x} 
        {\mu^2 \over m_{H^0}^2-m^2_{F_2} } \{
 -\cos(\alpha+\beta)[2c_\varphi+2xs_\varphi] \\
&&
 -\sin(\alpha+\beta)[{f\over h} (4x^2 c_\varphi + 6 c_\varphi +6xs_\varphi) 
                  + {A_h \over \mu} (c_\varphi +xs_\varphi)] \}
         \nonumber\\
\varepsilon_{A F'} &=& {v \over F\sub a} {x^2+3 \over x} 
        {\mu^2 \over m_{A}^2-m^2_{F'} } \{ {A_h\over \mu} -6{f\over h} \}
\end{eqnarray}
where $v \equiv \sqrt{\langle H_1\rangle^2+\langle H_2\rangle^2}$.
The flaton decay rates for the processes $F\to WW$ and $F'\to h^0Z$ of
 interest are given by
\begin{eqnarray} \label{DRates}
\Gamma(F_i \to WW) &=& \alpha_2 |\varepsilon_{VVF_i}|^2 
       {m_W^2 \over m_{F_i}}\sqrt{1-4{m_W^2\over m_{F_i}^2}}  \\
\Gamma(F' \to h^0Z) &\approx& {\alpha_2\over 8c_W^2}
  \cos(\alpha-\beta)^2 |\varepsilon_{AF'}|^2 m_{F'}
 \left(1+{m_{h^0}^2-{1\over2}m_Z^2 \over m_{F'}^2} \right) \nonumber
\end{eqnarray}
where $\varepsilon_{VVF_i} \equiv  \sin(\alpha-\beta)\varepsilon_{h^0F_i}
         - \cos(\alpha-\beta)\varepsilon_{H^0F_i}$.
The above rates have to be compared with the rates for the
decay processes $F\to aa$ and $F' \to aF_i$ [restricting ourselves
to the region (B) in Eq.~(\ref{AB})] which are given by \cite{denis}
\begin{eqnarray}
 \Gamma(F_1 \to aa) &=& {1\over 32\pi} {m_{F_1}^3\over F\sub a^2} 
                        {(-xs_\varphi+9c_\varphi)^2 \over (x^2+9)} 
           \nonumber\\
 \Gamma(F_2 \to aa) &=& {1\over 32\pi} {m_{F_2}^3\over F\sub a^2} 
                        {(xc_\varphi+9s_\varphi)^2 \over (x^2+9)} \\
 \Gamma(F' \to aF_i) &=& {1\over 16\pi} {m_{F'}^3\over F\sub a^2} 
                        \left(1-{m_{F_i}^2\over m_{F'}^2}\right)^3
                        {(3c_\varphi-3xs_\varphi)^2 \over (x^2+9)} 
\,.
   \nonumber
\end{eqnarray}
Let us 
define $R(F_i)\equiv \Gamma(F_i \to aa)/\Gamma(F_i \to WW)$ and 
$R(F')\equiv \Gamma(F' \to aF_i)/\Gamma(F' \to h^0Z)$ which behave like
\begin{equation} \label{Rs}
 R(F_i) \propto {m^4_{F_i} \over v^2 m_W^2} {m^4_{F_i} \over \mu^4}
 \,,\quad
 R(F') \propto {m^2_{F'} \over v^2} {(m_A^2-m^2_{F'})^2 \over \mu^4} 
\end{equation}
apart from the other numerical factors.
Eq.~(\ref{Rs}) shows that small $R$ can be easily obtained 
with small flaton masses and large $\mu$.  
In the case of small $\mu$, one would need to have small $m_A$ 
as well as $m_{F'}<v$.

Having calculated all of the relevant quantities, we could in principle
identify the region of parameter space which gives a viable cosmology
as well as viable particle physics. Here, we present instead
two  sets of parameters, with
 (i) a bino-like and
(ii) Higgsino-like  NLSP, which satisfy all the requirements.
The sets give an NLSP mass around the lowest possible value
$(m_{h^0}+m_Z)/2$.
The flatino LSP is only a little lighter, 
while the other flatino as well as the
flatons have masses around $200\GeV$, features which appear to be
typical for the region of parameter space corresponding to a light
NLSP.

In making the choice of parameters, we focussed on 
the region $ 2m_W < m_{F_1}, m_{F_2} < 2 m_Z$,
which can be arranged for $\xi \sim 12$ [see Eq.~(20)].
Both parameter sets have 
$$\tan\beta=3\,, \quad m_{h^0}=110 {\rm GeV}\,,\quad M_2=2M_1 \,,$$
and the other parameters are listed in
 in Table I along with the important output quantities.
In both cases, we have $R(F_i,F') < O(0.01)$ and 
the reheat temperature $T\sub{RH}=16$ GeV
using the decay rates (\ref{DRates}).  The  neutralino NLSP
 decouples after reheating, and can decay into
 the flatino LSP to  provide the
cold dark matter of the universe.  To check that potential overproduction of
the flatino LSP from 
the decay of thermal sfermions can be avoided, let us 
take as an example 
the right-handed stau.
{}From $\varepsilon_{\chi^0 \tilde F_1}$ in Table I, 
the coupling in Eq.~(\ref{17}) is given by  
$\varepsilon_{\tilde{\tau}_R \tau_R \tilde F_1} = 4.7\times10^{-8}$
which gives $\Omega_{\tilde F_1}h^2 \lesssim 1$ for
$m_{\tilde{\tau}_R} \gtrsim 300$ GeV for the case (i) and  a similar
figure is obtained for the case (ii).

\section{Conclusion}

In an interesting class of models, a non-renormalizable term of the
superpotential is responsible for the spontaneous breaking of 
PQ symmetry, while another such term generates the $\mu$ term
of the MSSM. The flaton fields which break PQ symmetry are
accompanied by flatinos, and the lightest flatino can be
the LSP. Through the $\mu$ term of the superpotential, the
NLSP neutralino decay might then be visible at colliders.

In this paper, we have examined the cosmology of this kind of model,
following our earlier work \cite{denis} which explored the same
flaton models on the assumption that all of the flatinos were unstable.
We make the reasonable assumption that the flaton fields are nonzero
in the early Universe, giving  thermal inflation which eliminates
pre-existing relics and leads to a rather well-defined cosmology.
The decay of flatons   into 
 relativistic axions will 
 interfere with nucleosynthesis unless the branching ratio is small,
while the decay of flatons into flatinos should
be forbidden altogether so as to avoid overproduction of the flatino
LSP. We have argued that these combined requirements lead to the
bounds  $m_{\tilde{\chi}_1^0} >
m_{\tilde F_1} > m_W$ or $ (m_Z +m_{h^0})/2$
on the masses of the neutralino NLSP $\tilde{\chi}_1^0$ and 
the flatino LSP $\tilde F_1$.
In order to have the lightest possible NLSP, one can 
focus on the case of light 
scalar (pseudoscalar) flatons, which can  decay only 
into two $W$ bosons (a light Higgs and a $Z$ boson).
These decays, as well as the neutralino NLSP decay into the 
flatino LSP,  come from the mixing between flatons (flatinos) and
Higgses (Higgsinos) due to the $\mu$ term of the superpotential.

We focus on the case of  a low axion scale, $F\sub a \sim 10^{10}$ GeV, 
which increases the feasibility of observing 
NLSP decays in future colliders. While the axion can  hardly
be the  dark matter with such a low scale,
the flatino LSP becomes a good dark matter
candidate, because  the reheat temperature can be high enough
to generate the NLSP in thermal equilibrium, but  low enough to suppress
flatino  production from the decay  of thermally produced sfermions. 

To make some definite statements, we studied
 an explicit model with $F\sub a\sim 10^{10}$ GeV. We verified 
 that
flaton decays into $WW$ or $h^0Z$ can be  large enough 
compared  to their decays into axions in a wide range of parameter space, 
 giving rise to the reheat temperature $T\sub{RH} \approx O(10)$ GeV.  
In this model,  the NLSP can be as light as $\sim 120\GeV$ 
with $m_{h^0}=110 $ GeV, while the other flatino and the flatons
have masses $\gsim 200\GeV$.  We further notice that
the parameter space admits the interesting 
mass region, $m_{\tilde{\chi}_1^0} -m_{\tilde F_1}=O(10)$ GeV, where
an enhancement of the  flaton--neutralino mixing,
$\varepsilon_{\tilde{\chi}_1^0 \tilde F_1} \sim 10^{-7}$, 
results and thus the NLSP decays becomes faster.


\begin{table}
\begin{center}
\begin{tabular}{|c|c|c|} \hline
		& Bino-like NLSP	& Higgsino-like NLSP	\\\hline\hline
\multicolumn{3}{|l|}{Inputs} \\\hline
$M_1$		& 120 GeV		& 220 GeV		\\\hline
$\mu$		& 287 GeV		& 130 GeV		\\\hline
$m_A$		& 300 GeV		& 150 GeV		\\\hline
$m_{H^0}$	& 305 GeV		& 162 GeV		\\\hline
$x$		& 4			& 4			\\\hline
$f/h$		& 1/24			& 1/11  		\\\hline
$A_h/\mu$	& -1			& -2.2			\\\hline
$A_f/\mu$	& 13/24			& 13/11  		\\\hline
\hline
\multicolumn{3}{|l|}{Outputs} \\\hline
$F\sub a$		& $6.6\times10^{10}$ GeV &$4.4\times10^{10}$ GeV\\\hline
$m_{F'}$	& 216 GeV		& 223 GeV		\\\hline
$m_{F_2}$	& 175 GeV		& 181 GeV		\\\hline
$m_{F_1}$	& 162 GeV		& 168 GeV		\\\hline
$m_{\tilde F_2}$& 184 GeV		& 190 GeV		\\\hline
$m_{\tilde F_1}$& 112 GeV		& 116 GeV		\\\hline
$m_{\tilde\chi^0_1}$& 121 GeV		& 123 GeV		\\\hline
$\epsilon_{\tilde\chi^0_1\tilde F_1}$
		& $-6.4\times10^{-8}$	& $37\times10^{-8}$	\\\hline
$\epsilon_{\chi^0\tilde F_1}$
  & $(15,-1.7,15.6,-4.8)v/F\sub a$ & $(10,-6.1,-44,-41)v/F\sub a$	\\\hline
$T\sub{RH}$	& 16 GeV		& 16 GeV		\\\hline
\end{tabular}
\end{center}
\caption{Two representative parameter sets
taking $\tan\beta=3$, $m_{h^0}=110$ GeV, $M_2=2M_1$ and $\xi=13$. 
Note that $\chi^0= (\tilde{B}, \tilde{W}_3, \tilde{H}_1^0, \tilde{H}_2^0)$
and $v=264$ GeV}
\end{table}

\end{document}